\documentclass[twocolumn,showpacs,preprintnumbers,amsmath,amssymb]{revtex4}

\usepackage{graphicx}
\usepackage{dcolumn}
\usepackage{bm}

\def\p{\partial}                                
\newcommand{\Aslash}{A\hspace{-2.0mm}/}         
\newcommand{\pslash}{\partial\hspace{-2.0mm}/}  

\begin{document}


\title{The absence of the 4$\psi$ divergence in noncommutative chiral models}

\author{Maja Buri\'c}%
\email{Maja.Buric@phy.bg.ac.yu}
\author{Du\v sko Latas}%
\email{Dusko.Latas@phy.bg.ac.yu}
\author{Voja Radovanovi\'c}%
\email{Voja.Radovanovic@phy.bg.ac.yu}
\affiliation{%
Faculty of Physics, University of Belgrade, \\ P.O.Box 368, RS-11001
Belgrade, Serbia
}%

\author{Josip Trampeti\'c}
\email{josipt@rex.irb.hr}
\affiliation{
Rudjer Bo\v skovi\' c Institute, Theoretical Physics Division, \\
P.O.Box 180 HR-10002 Zagreb, Croatia
}%

\date{\today}

\begin{abstract}
In this paper we show that in the noncommutative chiral gauge
theories the 4-fermion vertices are finite. The
$4\psi$-vertices appear in linear order in quantization of the
$\theta$-expanded noncommutative gauge theories; in all
previously considered models, based on Dirac fermions, the
$4\psi$-vertices were divergent and nonrenormalizable.
\end{abstract}

\maketitle

\section{\label{sec:Introduction}Introduction}

Although the issue of regularization of quantized field
theories was the original motivation to introduce the
noncommutativity of coordinates in the fifties~\cite{snyder},
the question of renormalizability of field theories on
noncommutative Minkowski space is still far from being settled.

In a definition of `noncommutative theory' there are several
steps which are not straightforward and need to be
specified. The first one is a definition of noncommutative
space. By noncommutative Minkowski space one usually means
the algebra of functions on  commutative ${\bf R}^4$ which
among themselves multiply with the Moyal-Weyl $\star$-product.
This product is associative but not commutative.
The $\star$-product of functions
integrated in the usual sense has a cyclic property, which is
necessary to define the action and with it the noncommutative
(NC) generalization of a classical field theory derived from
the action principle.

The second nonunique step is the very definition of a field theory.
This is because clearly, many theories can have the same commutative
limit. In the flat NC space a most straightforward way is to start
with a commutative theory and replace  the commuting fields with the
noncommuting ones and the ordinary products with the
$\star$-products. A noncommutative scalar $\phi^4$  theory and the
$\mathrm{U}(N)$ gauge theory have been formulated  in this manner
initially~\cite{phi4,nonab}. The most prominent result was that the
short and the long distances were related: this was seen through the
mixing of ultraviolet and infrared divergencies. The UV/IR mixing
also obstructs the renormalization. There are  other variants of
noncommutative scalar theories: as it has been shown
recently~\cite{grossephi4}, one can define a renormalizable
noncommutative  $\phi^4$ theory by modifying the original
commutative action by a potential $x^2\phi^2$. There are similar
proposals for  gauge theory, too~\cite{grossegauge}.

Generalization of the notion of gauge symmetry  is also not
unique. Inital proposals~\cite{nonab} contain UV/IR mixing as
does the scalar field theory. In the most recent models, the
symmetry in the ordinary sense is not changed; deformed is the
coproduct in the Hopf algebra~\cite{wessmarija}. These models
have not been tested for renormalizability yet. We shall work
in the framework of  the $\theta$-expanded gauge
theory~\cite{wess}. The original idea \cite{wulk} that, in
addition to gauge symmetry, nonuniqueness of the Seiberg-Witten
map  can be used to establish
renormalizability proved to be very useful in a couple of
models. A rough summary of the obtained
results~\cite{wulk,bichl,U1,SU2} is as follows. In general,
divergencies related to the gauge fields are weaker than those
for the fermions. When fermions are included an immediate
obstacle to renormalizability is found -- the so-called
`$4\psi$'-divergence, which is of the form
\begin{equation}
\mathcal{D}|_{\mathrm{div}}=\theta^{\mu\nu}
\epsilon_{\mu\nu\rho\sigma} (\bar\psi \gamma^\rho \psi) (\bar \psi
\gamma^\sigma \gamma_{5} \psi)~. \label{4psi}
\end{equation}
It appears independently of whether fermions are massive or
massless. This divergent vertex of the form of Fermi
interaction can not be regularized in any well-defined or
systematic way.

Recently, some potentially interesting and encouraging results on
renormalizability of the $\theta$-expanded theories have been
obtained~\cite{SUN,SM}. It was shown that the pure
$\mathrm{SU}(N)$ noncommutative gauge theory is renormalizable;
also, it was possible to define a generalization of the Standard
Model (SM) which has the gauge sector free of divergencies. These
results point out the importance of the choice of representation
for the renormalizability properties. With this in mind and recalling
that all previously considered models included only Dirac fermions,
we decided to redo the calculation of divergencies for the chiral
fermions. As an indicative and most important check we choose
to do first the $4\psi$-divergence. The result was: for chiral
fermions $4\psi$-divergence is absent! In this paper we present
the  calculations for the noncommutative $\mathrm{U}(1)$ and
$\mathrm{SU}(2)$ gauge theories.

\section{\label{sec:NCU(1)}Noncommutative U(1) theory}

\subsection{\label{Notation}Notation}

We will work in the noncommutative Minkowski space, defined by
the  relation
\begin{equation}
[x^\mu \stackrel{\star}{,} x^\nu] \equiv x^\mu \star x^\nu - x^\nu
\star x^\mu = \mathrm{i} \theta^{\mu\nu}~. \label{can}
\end{equation}
The commutator in~(\ref{can}) is a $\star$-commutator given by
the Moyal-Weyl product,
\begin{equation}
f(x)\star g(x) = \mathrm{e}^{\frac{\mathrm{i}}{2}
\theta^{\mu\nu}\frac{\partial}{\partial
x^\mu}\frac{\partial}{\partial y^\nu}}f(x)g(y)|_{y\to x}~.
\label{moyal}
\end{equation}

The action for the left chiral fermion $\varphi$ interacting
with the $\mathrm{U}(1)$ gauge field $A_\mu$ is, in commutative
theory, given by
\begin{eqnarray}
S_\mathrm{C} &=&\int \mathrm{d}^4x \;\mathcal{L}_{\mathrm{C}}
\nonumber \\
& = & \int \mathrm{d}^4x \left( \mathrm{i}\bar\varphi
\bar\sigma^\mu (\partial_\mu +\mathrm{i}A_\mu )\varphi -
\frac{1}{4}F_{\mu\nu}F^{\mu\nu} \right)~. \label{Scom}
\end{eqnarray}
The noncommutative $\mathrm{U}(1)$ symmetry can be realized by the
same set of fields. We denote the noncommutative gauge potential
by $\hat A_\mu $, the NC field strength by $\hat F_{\mu\nu} $,
$\hat F_{\mu\nu} = \partial_\mu \hat A_\nu - \partial_\nu \hat
A_\mu -[\hat A_\mu \stackrel{\star}{,} \hat A_\nu]$, and the NC Weyl
spinor by $\hat \varphi$. The noncommutative
$\mathrm{U}(1)$ symmetry is of course nonabelian; it can however
be related to the usual abelian $\mathrm{U}(1)$ symmetry by the
Seiberg-Witten map~\cite{Seiberg:1999vs}, which gives the basic,
the NC fields as  expansions in  their commutative
approximations. The SW map to first order in  $\theta^{\mu\nu}$
reads~\cite{wess,neutrino}
\begin{equation}
\hat A_\rho = A_\rho  -\frac 14 \theta^{\mu\nu} \{ A_\mu,
\partial_\nu A_\rho   + F_{\nu \rho} \}
+ \dots~, \label{expansion:A}
\end{equation}
\begin{eqnarray}
\hat F_{\rho\sigma} &= & F_{\rho\sigma}  +\frac{1}{2}\theta^{\mu\nu}
\{ F_{\mu\rho} ,F_{\nu\sigma} \} \label{expansion:F} \\
& & - \frac{1}{4}\theta^{\mu\nu}\{ A _\mu, (\partial_\nu + {D}_\nu
)F_{\rho\sigma}  \} +\dots~, \nonumber
\end{eqnarray}
\begin{equation}
\hat\varphi = \varphi
-\frac{1}{2}\theta^{\mu\nu} A_\mu \partial_\nu\varphi
+\frac{\mathrm{i}}{4}\theta^{\mu\nu}A_\mu A_\nu \varphi+
\dots~. \label{expansion:psi}
\end{equation}
The $\{\ ,\ \} $ denotes the anticommutator and ${D}_\mu$ is the
commutative covariant derivative. The fields $\varphi$ and strengths
$F_{\mu\nu} =\partial_\mu A_\nu - \partial_\nu A_\mu$
in ({\ref{expansion:A}-\ref{expansion:psi}) carry representations
of the commutative $\mathrm{U}(1)$ symmetry.

The action for noncommutative chiral electrodynamics is analogous
to~(\ref{Scom}):
\begin{eqnarray}
S_{\mathrm{NC}} & = &\int \mathrm{d}^4x \;\mathcal{L}_{\mathrm{NC}}
\nonumber \\
& = &
\int \mathrm{d}^4x \left( \mathrm{i}\hat{\bar \varphi}
\star \bar\sigma^\mu (\partial_\mu +\mathrm{i}\hat A_\mu ) \star
\varphi\phantom{\frac{1}{4}}\right. \\
& & \left.- \frac{1}{4}\hat F_{\mu\nu}\star \hat
F^{\mu\nu}\right)~. \nonumber
\end{eqnarray}
Expanding to first order in $\theta^{\mu\nu}$ we obtain
\begin{equation}
\mathcal{L}_{\mathrm{NC}} = \mathcal{L}_0 +\mathcal{L}_{1,A}
+\mathcal{L}_{1,\varphi}~, \label{lag}
\end{equation}
where
\begin{equation}
\mathcal{L}_0 = \mathcal{L}_{\mathrm{C}} =
\mathrm{i}\bar\varphi\sigma^\mu (\partial_\mu
+\mathrm{i} A_\mu )\varphi -
\frac{1}{4}F_{\mu\nu}F^{\mu\nu}~,
\label{L0}
\end{equation}
\begin{equation}
\mathcal{L}_{1,A} = -\frac{1}{2}\theta^{\mu\nu}
\left(F_{\mu\rho}F_{\nu\sigma}F^{\rho\sigma} -\frac{1}{4}
F_{\mu\nu}F_{\rho\sigma}F^{\rho\sigma}\right)~,
\label{L1A}
\end{equation}
\begin{equation}
\mathcal{L}_{1,\psi} =  -\frac{1}{8} \theta^{\mu\nu}
\Delta^{\alpha\beta\gamma}_{\mu\nu\rho}F_{\alpha\beta} \bar\varphi
\bar\sigma^\rho(\partial_\gamma
+\mathrm{i}A_\gamma)\varphi~.\label{L1varphi}
\end{equation}
The antisymmetric $\Delta$ is defined by
$\Delta^{\alpha\beta\gamma}_{\mu\nu\rho}=
-\epsilon^{\alpha\beta\gamma\delta}\epsilon_{\mu\nu\rho\delta}
$~. For nonabelian theories formulae~(\ref{L0}) and ~(\ref{L1A})
contain an additional trace in the group generators as we will
see in the special case of $\mathrm{SU}(2)$ later.

Obviously the parameter $\theta^{\mu\nu}$ of dimension
$(\mathrm{length})^2$ is small -- of order of magnitude $\lesssim
(\mathrm{TeV})^{-2}$ ~\cite{Z2Gamma}, and therefore the
expansion~(\ref{lag}) is useful to compute the almost-classical
effects of noncommutativity. Its relevance in considerations of
renormalizability is not quite clear since the divergent
contributions can be nonperturbative in $\theta^{\mu\nu}$; indeed
this is what happens with the UV/IR mixing. Nevertheless we shall
work with the truncated expression~(\ref{lag}) for two reasons:
first, an expansion like this might be a viable or a correct way
to define a renormalizable theory. Second, we presume that an
additional structure given by noncommutative Ward identities
exists: it then relates the $n$-point functions of different
orders in the $\theta$-expansion. Thu it is
possible, in principle, to use NC Ward identities in order to lift
renormalizability from $\theta$-linear to higher-$\theta$ orders.
In any case, if the theory is not renormalizable it will indeed show up
in linear order, which is also a result of relevance.

\subsection{\label{subsec:Quantization}Quantization}

We start with the action~(\ref{lag}) for noncommutative
chiral electrodynamics and quantize it by using the path-integral
method. We treat the $\theta$-dependent terms as interactions, the
parameter $\theta^{\mu\nu}$ as a coupling constant. The
propagators for the spinors and for the gauge fields are the same
as in the commutative theory. In order to compute the functional
integral we need either to complexify the gauge potential or to
work with the Majorana spinors; we choose the latter. Using
\begin{equation}
 \psi =
\left(
    \begin{array}{c}
        \varphi_\alpha \\
       \bar\varphi^{{\dot\alpha}}
    \end{array}
\right)~,
\end{equation}
we can rewrite the commutative part of the Lagrangian (\ref{Scom})
as
\begin{equation}
\mathcal{L}_0  =
\frac{\mathrm{i}}{2}\bar\psi\gamma^\mu(\partial_\mu
-\mathrm{i}\gamma_5 A_\mu )\psi -
\frac{1}{4}F_{\mu\nu}F^{\mu\nu}~. \label{L0'}
\end{equation}
To obtain~(\ref{L0'}) from (\ref{L0}) we use the $\gamma$-matrices
in the chiral representation; further details of the notation
are given in the Appendix. Written in terms of the Majorana spinors
the $\mathrm{U}(1)$ symmetry becomes axial; this is no suprise as
chiral lagrangian is not invariant under parity. For the
$\theta$-linear spinor part of the Lagrangian we obtain
\begin{equation}
\mathcal{L}_{1,\psi} =-\frac{1}{16} \theta^{\mu\nu}
\Delta^{\alpha\beta\gamma}_{\mu\nu\rho} F_{\alpha\beta}
\bar\psi\gamma^\rho(\partial_\gamma
-\mathrm{i}\gamma_5A_\gamma)\psi~,
\label{L1psi}
\end{equation}
while the gauge self-interaction is given by (\ref{L1A}).

In order to preserve gauge covariance we integrate using the
background field method~\cite{U1,SU2}. Briefly: in the first step we
expand fields around their classical configurations; we replace
therefore in the classical action
\begin{equation}
A_\mu\to  A_\mu + \mathcal{A}_\mu~,\qquad \psi \to \psi + \Psi~,
\end{equation}
where $A_\mu$, $\psi$ are the classical fields and
 $\mathcal{A}_\mu$, $\Psi$ the quantum fluctuations. After the
integration of the quantum fields, in the saddle point
approximation we obtain the one-loop  effective action
\begin{equation}
\Gamma[A_\mu,\psi] = S[A_\mu,\psi]
-\frac{1}{2\mathrm{i}}\mathrm{STr} \left(\log \mathcal{B}
[A_\mu,\psi]\right)~.
\label{*}
\end{equation}
The first term $ S[A_\mu,\psi]$ is the classical action and the
second is the one-loop quantum correction $\Gamma_1$. The operator
$\mathcal{B}[A_\mu,\psi] $, a result of Gaussian integration,
is the second functional derivative of $ S[A_\mu,\psi]$; it can be
obtained in our case by expanding $S[A_\mu+\mathcal{A}_\mu
,\psi+\Psi] $ to second order in $\mathcal{A}_\mu$ and $\Psi$
\begin{equation}
S^{(2)} = \int \mathrm{d}^4x
\begin{pmatrix}
    \mathcal{A}_\kappa & \bar\Psi
\end{pmatrix}
\mathcal{B}
\begin{pmatrix}
    \mathcal{A}_\lambda \\
    \Psi
\end{pmatrix}.
\end{equation}

We can divide $\mathcal{B}$ into its commutative part
$\mathcal{B}_0$ and a $\theta$-linear contribution
$\mathcal{B}_1$: $\mathcal{B} = \mathcal{B}_0 +\mathcal{B}_1$.
$\mathcal{B}_0$, after the gauge fixing, is given by
\begin{equation}
\mathcal{B}_0 = \frac{1}{2}
\begin{pmatrix}
    g^{\kappa\lambda}\Box & \bar\psi \gamma^\kappa \gamma_5 \\
    \gamma^\lambda \gamma_5 \psi & \mathrm{i}\pslash +\Aslash \gamma_5
\end{pmatrix}.
\label{B0}
\end{equation}
It contains the kinetic part
$$
\mathcal{B}_\mathrm{kin} =
 \frac{1}{2}
\begin{pmatrix}
    g^{\kappa\lambda}\Box &0 \\
   0 & \mathrm{i}\pslash
\end{pmatrix}~,
$$
and the interaction. In order to expand the logarithm in~(\ref{*})
around identity we have to multiply $\mathcal{B}$ by
$\mathcal{C}$,
\begin{equation}
\mathcal{C} = 2
\begin{pmatrix}
    g^{\kappa\lambda} & 0 \\
    0 & -\mathrm{i}\pslash
\end{pmatrix}.
\end{equation}
Then we can write
\begin{equation}
\mathcal{BC} = \Box \mathcal{I} +N_1 +T_1 +T_2~,
\label{3parts}
\end{equation}
with
\begin{equation}
\mathcal{I} =
\begin{pmatrix}
    g^{\kappa\lambda} & 0 \\
    0 & 1
\end{pmatrix}.
\end{equation}
The expression
\begin{eqnarray}
\Gamma_1 & =& \frac{\mathrm{i}}{2}
\mathrm{STr} \log \left(\mathcal{I}
+\Box^{-1}N_1 +\Box^{-1}T_1+\Box^{-1}T_2\right)\nonumber \\
&=&\frac{\mathrm{i}}{2} \sum\frac{(-1)^{n+1}}{n}
\mathrm{STr}\left(\Box^{-1}N_1+\Box^{-1}T_1\right.\label{perturb}
\\
& & \left.+\Box^{-1}T_2 \right)^n~, \nonumber
\end{eqnarray}
gives the perturbation expansion. $\Gamma_1$ can be identified
with the one-loop effective action because
\begin{equation}
\mathrm{STr} \left( \log \mathcal{B} \right)= \mathrm{STr}
\left(\log \Box^{-1}\mathcal{BC} \right) - \mathrm{STr}
\left(\mathcal{C} \Box^{-1}\right)~.
\end{equation}
As the last term does not depend on the fields $A_\mu$ and $\psi$ it
can be included in the (infinite) normalization.

In (\ref{3parts}) we have divided the interaction term in three
parts in the following way. Operator $N_1$ contains
the commutative 3-vertices, i.e. the terms with one classical
and two quantum fields. Analogously, operator $T_1$ is a term linear in
$\theta^{\mu\nu}$ containing one classical field, and $T_2$ is
linear in $\theta^{\mu\nu}$ containing two classical
fields. From~(\ref{B0}) we see that $N_1$ equals to
\begin{equation}
N_1 =
\begin{pmatrix}
    0 & -\mathrm{i}\bar\psi\gamma^\kappa\gamma_5 \pslash \\
    \gamma^\lambda\gamma_5\psi & \gamma_5\Aslash \pslash
\end{pmatrix}.
\end{equation}
The noncommutative vertices $T_1$ and $T_2$ require a bit more
work. Using the Majorana spinor-identities we obtain
\begin{widetext}
\begin{equation}
T_1 = \frac{1}{8} \theta^{\mu\nu}\Delta^{\alpha\beta\gamma}_{\mu\nu\rho}
\begin{pmatrix}
    V^{\rho\kappa\lambda}_{\alpha\beta\gamma}  &
    -2 \delta^\kappa_\alpha (\partial_\beta\bar\psi)\gamma^\rho
    \partial_\gamma \pslash \\
    - 2 \mathrm{i} \delta^\lambda_\alpha \gamma^\rho (\partial_\beta \psi)
    \partial_\gamma &
    - F_{\alpha\beta}\gamma^\rho \partial_\gamma \pslash
\end{pmatrix},
\end{equation}
and
\begin{equation}
T_2 = \frac{1}{8}
\theta^{\mu\nu}\Delta^{\alpha\beta\gamma}_{\mu\nu\rho}
\begin{pmatrix}
    \delta^\kappa_\alpha \delta^\lambda_\beta
    (2\bar\psi\gamma^\rho\gamma_5 \psi\partial_\gamma
    +(\partial_\gamma \bar\psi \gamma^\rho \gamma_5 \psi) ) &
    2 \mathrm{i} \delta^\kappa_\alpha( (\partial_\beta \bar\psi)
    \gamma^\rho\gamma_5 A_\gamma + \bar\psi\gamma^\rho \gamma_5
    F_{\beta\gamma} -\bar\psi\gamma^\rho \gamma_5 A_\beta \partial_\gamma)
    \pslash \\
    -\delta^\lambda_\alpha (2\gamma^\rho \gamma_5 A_\beta\psi\partial_\gamma
    +F_{\beta\gamma}\gamma^\rho \gamma_5\psi ) &
    \mathrm{i} A_\alpha F_{\beta\gamma}\gamma^\rho\gamma_5 \pslash
\end{pmatrix}, \nonumber
\end{equation}
\end{widetext}
with the bosonic part $V^{\rho\kappa\lambda}_{\alpha\beta\gamma}$
given in~\cite{U1}. As we are looking just the four-fermion
divergence, $V^{\rho\kappa\lambda}_{\alpha\beta\gamma}$ will not
contribute to our calculations as we shall shortly explain.

\subsection{\label{subsec:Divergencies}Divergencies}

We have noted, the full perturbation expansion is given
in (\ref{perturb}). For example: the 4-point functions are contained
in terms which have the sum of the operator-indices equal to 4. If in
addition we look only for the contributions linear in
$\theta^{\mu\nu}$, the relevant expressions can have at most one
of the $T_1$ or $T_2$. There are two such terms,
\begin{equation*}
\mathcal{D}_1 =\mathrm{STr}\left((\Box ^{-1}N_1)^3 (\Box ^{-1}T_1)
\right)~,
\end{equation*}
and
\begin{equation*}
\mathcal{D}_2 =\mathrm{STr} \left((\Box ^{-1}N_1)^2 (\Box ^{-1}T_2)
\right)~.
\end{equation*}
As here we restrict to the 4-fermion vertex, we can further simplify
the calculation by putting  $A_\mu =0$. Under this condition we also
have $V^{\rho\kappa\lambda}_{\alpha\beta\gamma} =0$.

In order to find the divergencies we write the traces in the
momentum representation and afterwards perform the dimensional
regularization. The result which we obtain is: the term
$\mathcal{D}_1$ is finite. For the divergent part of $\mathcal{D}_2$ we
get
\begin{equation}
\mathcal{D}_2\vert_{\mathrm{div}} =\frac{1}{(4\pi )^2\epsilon}
\frac{3\mathrm{i}}{8}\epsilon_{\mu\nu\rho\sigma}\theta^{\mu\nu}
(\bar\psi\gamma^\rho\gamma_5\psi)
(\bar\psi\gamma^\sigma\gamma_5\psi)~.
\end{equation}
However the last expression vanishes too, due to the antisymmetry
of the Levi-Civita symbol. In fact, in
retrospective, it is easy to see that the $4\psi$-divergence has to be
zero in the chiral case: because of antisymmetry of
$\epsilon^{\mu\nu\rho\sigma}$ the only possible expression
is (\ref{4psi}). On the other hand  for the Majorana spinors
$\bar\psi\gamma^\mu\psi \equiv 0$ and therefore
(\ref{4psi}) vanishes identically.

\section{\label{sec:NCSU(2)}Noncommutative SU(2) theory}

\subsection{\label{subsec:Lagrangian}Lagrangian}
We will now do an analogous analysis for the chiral fermions in
the fundamental representation of $\mathrm{SU}(2)$.
We start with a
doublet of fermions and a vector potential:
\begin{equation}
\varphi =
\begin{pmatrix}
    \varphi_1 \\ \varphi_2
\end{pmatrix},
\quad A_\mu =A_\mu^a \frac{\sigma_a}{2}~,
\end{equation}
and the following commutative Lagrangian
\begin{eqnarray}
\mathcal{L}_{0} &=& \mathcal{L}_{0,\psi} +\mathcal{L}_{0,A} \nonumber \\
& = & \mathrm{i}\bar\varphi \bar\sigma^\mu (\partial_\mu +\mathrm{i}
A_\mu ) \varphi- \frac{1}{4}\mathrm{Tr} F_{\mu\nu}F^{\mu\nu}~,
\end{eqnarray}
which we want to rewrite in terms of the Majorana spinors
\begin{equation}
\psi_{1} =
\begin{pmatrix}
    \varphi_1 \\ \bar\varphi_1
\end{pmatrix},
\quad
\psi_{2} =
\begin{pmatrix}
    \varphi_2 \\ \bar\varphi_2
\end{pmatrix}~.
\end{equation}
As the fundamental representation of $\mathrm{SU}(2)$ is not real, when
we write the Lagrangian in the Majorana spinors we apparently break
the $\mathrm{SU}(2)$ symmetry, i.e. we have to write each
component of the vector potential separately. That is
\begin{eqnarray}
\mathcal{L}_{0, \psi} &=& \mathrm{i}\bar\varphi_1\bar
\sigma^\mu(\partial_\mu +\frac{\mathrm{i}}{2} A_\mu^3) \varphi_1
-\frac{1}{2} \bar\varphi_1\bar
\sigma^\mu A_\mu^-\varphi_2 \nonumber\\
& &-\frac{1}{2}\bar\varphi_2\bar \sigma^\mu A_\mu^+\varphi_1
+\mathrm{i}\bar\varphi_2\bar\sigma^\mu(\partial_\mu
-\frac{\mathrm{i}}{2} A_\mu^3)\varphi_2 \nonumber\\
& = &\frac{1}{4}
\begin{pmatrix}
    \bar\psi_1 &\bar\psi_2
\end{pmatrix}
\begin{pmatrix}
    2\mathrm{i}\pslash+\Aslash_3 \gamma_5  & \Aslash_1 \gamma_5
    +\mathrm{i}\Aslash_2 \\[5pt]
    \Aslash_1 \gamma_5 - \mathrm{i} \Aslash_2 &
    2 \mathrm{i}\pslash-\Aslash_3 \gamma_5
\end{pmatrix}
\begin{pmatrix}
    \psi_1 \\ \psi_2
\end{pmatrix}~. \nonumber
\end{eqnarray}
We have denoted $A_\mu^\pm = A_\mu^1 \pm \mathrm{i}A_\mu^2 $. Now of
course $ \begin{pmatrix}\psi_1 \\ \psi_2
\end{pmatrix} $ is not a $\mathrm{SU}(2)$ doublet.

The $\theta$-linear bosonic part of the $\mathrm{SU}(2)$ Lagrangian
\begin{eqnarray}
\mathcal{L}_{1,A} &=& -\frac{1}{2}\ \theta^{\rho\sigma}
\mathrm{Tr}\,(F_{\mu\rho}F_{\nu\sigma} -\frac{1}{4}
F_{\rho\sigma}F_{\mu\nu})F^{\mu\nu}\nonumber\\
&=&  -\frac{1}{2}\ d^{abc}\theta^{\rho\sigma} (F_{\mu\rho}^a
F_{\nu\sigma}^b  -\frac{1}{4} F_{\rho\sigma}^aF_{\mu\nu}^b)
F^{\mu\nu c}~, \nonumber
\end{eqnarray}
vanishes, because it is proportional to the symmetric
coefficients $d^{abc}$
\begin{equation}
d^{abc}\sim \mathrm{Tr}\left(\sigma^a\{ \sigma^b, \sigma^c \}\right)
= 0~.
\end{equation}
In fact, $d^{abc} =0 $ for all irreducible representations of
$\mathrm{SU}(2)$.  On the other hand, the fermionic linear
part of the Lagrangian,
\begin{widetext}
\begin{equation}
\mathcal{L}_{1,\psi} = \frac{1}{32}\theta^{\mu\nu}
\Delta^{\alpha\beta\gamma}_{\mu\nu\rho}
\bar\varphi
\bar\sigma^\rho \left(2 \mathrm{i} F_{\alpha\beta}^a \sigma^a
\partial_\gamma + A_\alpha^a F_{\beta\gamma}^a -\mathrm{i}
\epsilon^{abc} A_\alpha^a F_{\beta\gamma}^b\sigma^c\right) \varphi~,
\end{equation}
in the Majorana representation is rewritten as
\begin{equation}
\mathcal{L}_{1,\psi} = \frac{1}{64} \theta^{\mu\nu}
\Delta^{\alpha\beta\gamma}_{\mu\nu\rho}
\begin{pmatrix}
    \bar\psi_1 & \bar\psi_2
\end{pmatrix}
\begin{pmatrix}
    2\mathrm{i}F_{\alpha\beta}^3 \gamma^\rho \p_\gamma
    + A_\alpha ^a F_{\beta\gamma}^a \gamma^\rho \gamma_5  &
    2 \mathrm{i} F_{\alpha\beta}^1 \gamma^\rho \p_\gamma
    - 2 F_{\alpha\beta}^2 \gamma^\rho\gamma_5\p_\gamma \\
    2 \mathrm{i}  F_{\alpha\beta}^1\gamma^\rho \p_\gamma
    +2 F_{\alpha\beta}^2\gamma^\rho\gamma_5\p_\gamma &
    -2 \mathrm{i} F_{\alpha\beta}^3 \gamma^\rho \p_\gamma
    + A_\alpha ^a F_{\beta\gamma}^a \gamma^\rho \gamma_5
\end{pmatrix}
\begin{pmatrix}
    \psi_1 \\ \psi_2
\end{pmatrix}.
\nonumber
\end{equation}
\end{widetext}

\subsection{\label{subsec:NCSU(2)QandD}Quantization and divergencies}

As we have written the Lagrangian in an appropriate form, the
procedure of quantization is straightforward and follows closely
that which was done for the $\mathrm{U}(1)$. It is interesting that
the results, as we shall see shortly, are completely analogous,
though the intermediate calculations are now considerably more
involved. The part of the action which is of second order in quantum
fields we write as
\begin{equation}
S^{(2)} =\int \mathrm{d}^4 x
\begin{pmatrix}
    \mathcal{A}_\zeta^1 & \mathcal{A}_\eta^2 & \mathcal{A}_\xi ^3 &
    \bar\Psi_1 & \bar\Psi_2
\end{pmatrix}
\mathcal{B}
\begin{pmatrix}
    \mathcal{A}_{\zeta^\prime}^1 \\
    \mathcal{A}_{\eta^\prime}^2 \\
    \mathcal{A}_{\xi^\prime}^3 \\
    \Psi_1 \\
    \Psi_2
\end{pmatrix}.
\end{equation}term
Matrices $\mathcal{B}_\mathrm{kin}$ and $\mathcal{C}$ have the same,
just enlarged, structure as in the  $\mathrm{U}(1)$ case:
\begin{eqnarray}
\mathcal{B}_\mathrm{kin} &=& \frac{1}{2} \left(
    \begin{array}{c|c}
        \begin{array}{ccc}
        \Box\delta^{\zeta {\zeta^\prime}} &  &  \\
         &\Box\delta^{\eta {\eta^\prime}}  &  \\
         & &\Box\delta^{\xi {\xi^\prime}}
    \end{array}
    & 0 \\ \hline
    0 &
        \begin{array}{cc}
        \mathrm{i}\pslash &  \\
         & \mathrm{i}\pslash
        \end{array}
    \end{array}
\right)\nonumber \\ & = & \frac{1}{2} \mathrm{diag} \left(
\Box\delta^{\zeta {\zeta^\prime}} \quad \Box\delta^{\eta
{\eta^\prime}}\quad \Box\delta^{\xi {\xi^\prime}} \quad
\mathrm{i}\pslash \quad \mathrm{i}\pslash \right), \nonumber
\end{eqnarray}
\begin{eqnarray}
\mathcal{C} & = & 2 \left(
   \begin{array}{c|c}
   \begin{array}{ccc}
   \delta^{\zeta {\zeta^\prime}} & & \\
    & \delta^{\eta {\eta^\prime}} & \\
    & & \delta^{\xi {\xi^\prime}}
   \end{array} & 0 \\ \hline
   0 &
        \begin{array}{cc}
        -\mathrm{i}\pslash &  \\
         & -\mathrm{i}\pslash
        \end{array}
   \end{array}
\right) \nonumber \\
& = & 2 \mathrm{diag} \left( \delta^{\zeta {\zeta^\prime}} \quad
\delta^{\eta {\eta^\prime}} \quad \delta^{\xi {\xi^\prime}} \quad
-\mathrm{i}\pslash \quad -\mathrm{i}\pslash \right).\nonumber
\end{eqnarray}
The interactions of course differ: now we have the 4-boson vertex in
the commutative part for example, etc. The one-loop effective action
has the form
\begin{equation}
\Gamma_1 =\frac{\mathrm{i}}{2} \mathrm{STr}\,
\log \left(\mathcal{I} +\Box^{-1}(N_1 +N_2+T_1+T_2+T_3)\right).
\end{equation}

As the formulae are cumbersome we shall here restrict
immediately to the subset of fermionic diagrams defined by the
condition $A_\mu^a =0$, which of course simplifies  the
calculations. With this restriction we also have $N_2
=0$ and $T_3 =0$. The remaining interaction vertex is
\begin{widetext}
\begin{equation}
N_1 = \frac{1}{2}
\left(
    \begin{array}{c|c}
        0 & \begin{array}{cc}
        -\mathrm{i} \bar\psi_2\gamma^\zeta \gamma_5 \pslash &
        -\mathrm{i}\bar\psi_1\gamma^\zeta \gamma_5 \pslash\p \\
       - \bar\psi_2\gamma^\eta \pslash &
        \bar\psi_1\gamma^\eta \pslash \\
        -\mathrm{i}\bar\psi_1\gamma^\xi \gamma_5 \pslash &
        \mathrm{i}\bar\psi_2\gamma^\xi \gamma_5 \pslash
            \end{array} \\ \hline
        \begin{array}{ccc}
        \gamma^{\zeta^\prime}\gamma_5\psi_2 &
        \mathrm{i}\gamma^{\eta^\prime}\psi_2 &
        \gamma^{\xi^\prime}\gamma_5\psi_1  \\
        \gamma^{\zeta^\prime}\gamma_5\psi_1 &
       - \mathrm{i}\gamma^{\eta^\prime}\psi_1 &
        -\gamma^{\xi^\prime}\gamma_5\psi_2
    \end{array} & 0
    \end{array}
\right)~.
\end{equation}
In $\theta$-linear order we have the following $T_1$ matrix:
\begin{eqnarray}
T_1 &=& -\frac{1}{8}\theta^{\mu\nu}
\Delta^{\alpha\beta\gamma}_{\mu\nu\rho} \nonumber \\
& & \times\left(
    \begin{array}{c|c}
        0 & \begin{array}{cc}
        \delta^\zeta_\alpha (\partial_\beta \bar\psi_2)
        \gamma^\rho \partial_\gamma\pslash &
        \delta^\zeta_\alpha (\partial_\beta \bar\psi_1)
        \gamma^\rho \partial_\gamma \pslash \\
       - \mathrm{i}\delta^\eta_\alpha (\partial_\beta \bar\psi_2)
        \gamma^\rho \gamma_5\partial_\gamma \pslash &
        \mathrm{i}\delta^\eta_\alpha (\partial_\beta \bar\psi_1)
        \gamma^\rho \gamma_5\partial_\gamma \pslash \\
        \delta^\xi_\alpha(\partial_\beta \bar\psi_1)
        \gamma^\rho \partial_\gamma \pslash &
        -\delta^\xi_\alpha(\partial_\beta \bar\psi_2)
        \gamma^\rho \partial_\gamma \pslash
            \end{array} \\ \hline
        \begin{array}{ccc}
        \mathrm{i}\delta^{\zeta^\prime}_\alpha
        \gamma^\rho (\partial_\beta \psi_2)\partial_\gamma &
        -\delta^{\eta^\prime}_\alpha \gamma^\rho \gamma_5
        (\partial_\beta\psi_2)\partial_\gamma &
        \mathrm{i}\delta^{\xi^\prime}_\alpha \gamma^\rho
        (\partial_\beta \psi_1)\partial_\gamma  \\
        \mathrm{i}\delta^{\zeta^\prime}_\alpha \gamma^\rho
        (\partial_\beta \psi_1)\partial_\gamma &
        \delta^{\eta^\prime}_\alpha \gamma^\rho \gamma_5
        (\partial_\beta \psi_1)\partial_\gamma &
        -\mathrm{i}\delta^{\xi^\prime}_\alpha \gamma^\rho
        (\partial_\beta\psi_2)\partial_\gamma
        \end{array} & 0
    \end{array}
\right)~. \nonumber
\end{eqnarray}
while, $T_2$ matrix is given by
\begin{equation}
T_2 =
\frac{1}{32}\,\theta^{\mu\nu}\Delta^{\alpha\beta\gamma}_{\mu\nu\rho}
\left(
    \begin{array}{c|c}
        \begin{array}{ccc}
        2 \delta^\zeta_\alpha \delta^{\zeta^\prime}_\beta E^\rho \partial_\gamma &
        \delta^\eta_\alpha \delta^{\zeta^\prime}_\beta F^\rho_\gamma &
        - \delta^\xi_\alpha \delta^{\zeta^\prime}_\beta  G^\rho_\gamma \\
        - \delta^\zeta_\alpha \delta^{\eta^\prime}_\beta F^\rho_\gamma &
        2 \delta^\eta_\alpha \delta^{\eta^\prime}_\beta  E^\rho \partial_\gamma &
        \delta^\xi_\alpha \delta^{\zeta^\prime}_\beta H^\rho_\gamma \\
        \delta^\zeta_\alpha \delta^{\xi^\prime}_\beta G^\rho_\gamma &
        -\delta^\eta_\alpha \delta^{\xi^\prime}_\beta H^\rho_\gamma &
        2 \delta^\xi_\alpha \delta^{\xi^\prime}_\beta  E^\rho \partial_\gamma
        \end{array} & 0 \\ \hline
        0 & 0
    \end{array}
\right)~,
\end{equation}
\end{widetext}
where
\begin{equation}
E^\rho =\bar\psi_1\gamma^\rho\gamma_5\psi_1
+\bar\psi_2\gamma^\rho\gamma_5\psi_2~,
\end{equation}
\begin{eqnarray}
F^\rho_\gamma &=& -\mathrm{i} (\partial_\gamma
\bar\psi_1)\gamma^\rho\psi_1
+\mathrm{i} \bar\psi_1\gamma^\rho (\partial_\gamma\psi_1) \nonumber \\
& & +\mathrm{i}(\partial_\gamma\bar\psi_2)\gamma^\rho\psi_2
-\mathrm{i}\bar\psi_2\gamma^\rho(\partial_\gamma \psi_2)~,
\end{eqnarray}
\begin{eqnarray}
G^\rho_\gamma &=& (\partial_\gamma \bar\psi_1)\gamma^\rho
\gamma_5\psi_2
- \bar\psi_1\gamma^\rho \gamma_5(\partial_\gamma \psi_2) \nonumber \\
& & -(\partial_\gamma \bar\psi_2)\gamma^\rho \gamma_5\psi_1
+\bar\psi_2\gamma^\rho \gamma_5(\partial_\gamma \psi_1)~,
\end{eqnarray}
\begin{eqnarray}
H^\rho_\gamma &=&-\mathrm{i} (\partial_\gamma
\bar\psi_1)\gamma^\rho\psi_2
+\mathrm{i} \bar\psi_1\gamma^\rho (\partial_\gamma \psi_2) \nonumber \\
& & -\mathrm{i}(\partial_\gamma \bar\psi_2)\gamma^\rho \psi_1
+\mathrm{i}\bar\psi_2\gamma^\rho(\partial_\gamma \psi_1)~.
\end{eqnarray}

As before, the divergent contributions in principle come from
$\mathcal{D}_1$ and $\mathcal{D}_2$. However, $\mathcal{D}_1$ is
finite, while for the divergent part of $\mathcal{D}_2$ we obtain
\begin{eqnarray}
\mathcal{D}_2\vert_{\mathrm{div}} &=& -\frac{1}{(4\pi)^2\epsilon}\,
\frac{9\mathrm{i}}{64} \,\theta^{\mu\nu} \epsilon_{\mu\nu\rho\sigma}
(\bar\psi_1\gamma^\rho\gamma_5\psi_1
+\bar\psi_2\gamma^\rho\gamma_5\psi_2)
\nonumber \\
& & \times (\bar\psi_1\gamma^\sigma \gamma_5\psi_1 +
\bar\psi_2\gamma^\sigma \gamma_5\psi_2)~,
\end{eqnarray}
which identically vanishes, too.

\section{\label{sec:Anomalies} Conclusions}
When one thinks  about the quantization of the chiral models, the first
question which naturally arises is the one about anomalies. 
The issue of chiral anomalies for the $\theta$-expanded models has
been analyzed in details in~\cite{anomaly}, and the result was
that, for the compact gauge groups, anomalies are the same as in the
commutative theory. This, for example, means that the noncommutative
chiral electrodynamics which we analyzed in Section~2 cannot be
quantized consistently. But on the other hand, it also means
that we can build the particle physics models as in the
ordinary theory, for example the noncommutative chiral
$\mathrm{U}(1)\times \mathrm{SU}(2)$ gauge theory is consistent if
lepton and quark multiplets are the same as in the Standard Model.
Our present result asserts that in addition such  model has no
four-fermion divergencies.

Construction of a consistent
noncommutative standard model (NCSM is  the main motivation of our
investigation.  We have previously  proposed a model which is
renormalizable in the gauge sector~\cite{SM,Z2Gamma} and the present
result opens a possibility to extend it. Of course the Higgs sector
should also be investigated~\cite{higgs}. There is a number of
phenomenological predictions of the NCSM models
\cite{Goran,Blazenka,Josip,Ohl:2004tn}; they would become more
robust if one could prove the one-loop renormalizability.

Obviously,  there is  a long way to go to show the full
renormalizability of the NC chiral gauge models: what we have done
here is just an initial step. As from the one-loop renormalizability
no immediate conclusions can be made about the all-loop properties,
likewise from the renormalizability in $\theta$-linear order nothing
automatically follows for the full SW expansion. There are many
steps to be done: some, as extension from linear to higher orders in
$\theta^{\mu\nu}$, we just see as viable possibilities. Other, like
the analysis of all one-loop divergent vertices in the $\theta$ 
linear order, are
straightforward and require additional work, and this is what we
plan to do in our following work. One should remember that in the
$\theta$-expanded theories one has an additional tool for
renormalizability, the SW field redefinition. Note also that the
renormalizability principle could help to minimize or even cancel
most of the ambiguities of the higher order SW
maps~\cite{Moller:2004qq}.

If indeed the $\theta$-linear order of the chiral gauge models
proves to be renormalizable, then it will really be important to
analyze the noncommutative Ward identities and their
implications to renormalizability more systematically.

\begin{acknowledgments}
The work of M.~B., V.~R. and D.~L. is a done within the project
141036 of the Serbian Ministry of Science. The work of J.~T. is
supported by the project 098-0982930-2900 of the Croatian Ministry
of Science Education and Sports. Our collaboration was partly
supported by the UNESCO project 875.834.6 through the SEENET-MTP and
by ESF in the framework of the Research
Networking Programme on 'Quantum Geometry and Quantum Gravity'.
\end{acknowledgments}

\appendix

\section{\label{App:sec:Conv}Conventions}

The notation and the rules of  chiral-spinor algebra follow
basically~\cite{chiral}. We use the following chiral representation
of the $\gamma$-matrices
\begin{equation}
\gamma^\mu =
\begin{pmatrix}
0 & \sigma^\mu \\
\bar\sigma^\mu & 0
\end{pmatrix}~,
\quad
\gamma_5 =
\begin{pmatrix}
-1 & 0\\
0 & 1
\end{pmatrix}~,
\end{equation}
with
\begin{equation}
\sigma^\mu = (1,\vec\sigma)~, \quad \bar\sigma^\mu = (1,
-\vec\sigma)~.
\end{equation}
This means in particular
$$
\bar\sigma^{\mu{\dot\alpha}\alpha}
=\epsilon^{{\dot\alpha}{\dot\beta}}
\epsilon^{\alpha\beta}\sigma^\mu_{\beta{\dot\beta}}~.
$$
The chiral $\psi$, $\chi$ spinors multiply as
\begin{equation}
\varphi \chi =\chi\varphi~,\quad \bar\varphi \bar\chi
=\bar\chi\bar\varphi~, \label{ind}
\end{equation}
\begin{equation}
\bar\varphi\bar\sigma^\mu \chi =-\chi\sigma^\mu\bar\varphi~, \quad
(\chi\sigma^\mu\bar\varphi )^\dagger =\varphi\sigma^\mu\bar\chi~.
\nonumber
\end{equation}
Those relations, as can be seen easily, give the usual identities
for the Majorana spinors $\phi$, $\psi$ which we use
\begin{equation}
\bar\phi \psi =\bar\psi \phi~,\quad \bar\phi\gamma_5 \psi =\bar\psi
\gamma_5\phi~,\nonumber
\end{equation}
\begin{equation}
\bar\phi \gamma^\mu \psi = -\bar\psi \gamma^\mu \phi~,\quad \bar\phi
\gamma^\mu\gamma_5\psi =\bar\psi\gamma^\mu\gamma_5 \phi~. \nonumber
\end{equation}
Majorana Lagrangians are obtained from the corresponding chiral
ones using the identities~(\ref{ind}) and the fact that  Lagrangians
are real.

\end{document}